\documentclass{NScrust}

\begin{document}

\markboth{Jun Xu et al.}{Neutron star crust}

\catchline{}{}{}{}{}

\title{TRANSITION DENSITY AND PRESSURE AT THE INNER EDGE OF NEUTRON STAR CRUSTS}

\author{JUN XU and CHE MING KO}

\address{Cyclotron Institute and Department of Physics and Astronomy, Texas A\&M University,
College Station, TX 77843-3366, USA\\
xujun@comp.tamu.edu and ko@comp.tamu.edu}

\author{LIE-WEN CHEN}

\address{Department of Physics, Shanghai Jiao Tong University, Shanghai
200240, China\\
Center of Theoretical Nuclear Physics, National Laboratory of
Heavy-Ion Accelerator, Lanzhou, 730000, China\\
lwchen@sjtu.edu.cn}

\author{BAO-AN LI}

\address{Department of Physics and Astronomy, Texas A\&M University-Commerce, Commerce, TX
75429-3011, USA\\
Bao-An\_Li@tamu-commerce.edu}

\author{HONG-RU MA}

\address{Department of Physics, Shanghai Jiao Tong University, Shanghai
200240, China\\
hrma@sjtu.edu.cn}

\maketitle

\begin{history}
\received{(received date)}
\revised{(revised date)}
\end{history}

\begin{abstract}

Using the nuclear symmetry energy that has been recently constrained
by the isospin diffusion data in intermediate-energy heavy ion
collisions, we have studied the transition density and pressure at
the inner edge of neutron star crusts, and they are found to be
$0.040$ fm$^{-3}$ $\leq \rho _{t}\leq 0.065$ fm$^{-3}$ and $0.01$
MeV/fm$^{3}$ $\leq P_{t}\leq 0.26$ MeV/fm$^{3}$, respectively, in
both the dynamical and thermodynamical approaches. We have also
found that the widely used parabolic approximation to the equation
of state of asymmetric nuclear matter gives significantly higher
values of core-crust transition density and pressure, especially for
stiff symmetry energies. With these newly determined transition
density and pressure, we have obtained an improved relation between
the mass and radius of neutron stars.

\end{abstract}

\section{Introduction}
\label{introduction}

Studying the properties of  neutron stars, which are among the most
mysterious objects in the universe, allows us to test our knowledge
of matter under extreme conditions. Theoretical studies have shown
that neutron stars are expected to have a liquid core surrounded by
a solid inner crust~\cite{Cha08}, which extends outward to the
neutron drip-out region. While the neutron drip-out density $\rho
_{\rm out}$ is relatively well determined to be about $4\times
10^{11}$ g/cm$^{3}$ or $0.00024~{\rm fm}^{-3}$~\cite{Rus06}, the
transition density $\rho _{t}$ at the inner edge is still largely
uncertain mainly because of our very limited knowledge on the
nuclear equation of state (EOS), especially the density dependence
of the symmetry energy ($E_{\rm sym}(\rho)$) of neutron-rich nuclear
matter~\cite{Lat00,Lat07}. These uncertainties have hampered our
understanding of many important properties of neutron
stars~\cite{Lat00,Lat07,Lat04} and related astrophysical
observations~\cite{Lat00,Lat07,BPS71,BBP71,Pet95a,Pet95b,Ste05,Lin99,Hor04,Bur06,Owe05}.
Recently, significant progress has been made in constraining the EOS
of neutron-rich nuclear matter using terrestrial laboratory
experiments (See Ref.~\cite{LCK08} for the most recent review). In
particular, the analysis of the isospin-diffusion data
\cite{Tsa04,Che05a,LiBA05c} in heavy-ion collisions has constrained
tightly the $E_{\rm sym}(\rho)$ in exactly the same sub-saturation
density region around the expected inner edge of neutron star crust.
The extracted slope parameter $L=3\rho_0\frac{\partial E_{\rm
sym}(\rho)}{\partial\rho}|_{\rho=\rho_0}$ in the density dependence
of the nuclear symmetry energy was found to be $86\pm25$ MeV, which
has further been confirmed by a more recent analysis~\cite{Tsa09}.
With this constrained nuclear symmetry energy, we have obtained an
improved determination of the values for the transition density and
pressure at the inner edge of neutron star crusts. This has led us
to obtain significantly different values for the radius of the Vela
pulsar from those estimated previously. Also, we have found that the
widely used parabolic approximation (PA) to the EOS of asymmetric
nuclear matter gives much larger values for the transition density
and pressure, especially for stiff symmetry energies.

\section{Dynamical and Thermodynamical approaches to the stability of $npe$ matter}
\label{approaches}

The transition density at the inner edge of neutron star crusts can
be determined using the
dynamical~\cite{BPS71,BBP71,Pet95a,Pet95b,Dou00,Oya07,Duc07} and the
thermodynamical approaches~\cite{Lat07,Kub07,Wor08} as well as the
approach based on the random phase approximation
(RPA)~\cite{Hor01,Hor03}. Our study was based on the dynamical and
thermodynamical approaches.

In the dynamical approach, the stability condition for a homogeneous
$npe$ matter against small periodic density perturbations can be well approximated by \cite%
{BPS71,BBP71,Pet95a,Pet95b,Oya07}
\begin{equation}
V_{\rm dyn}(k)=V_{0}+\beta k^{2}+\frac{4\pi e^{2}}{k^{2}+k_{TF}^{2}}>0,
\label{Vdyn}
\end{equation}%
where $k$ is the wavevector of the spatially periodic density
perturbations and
\begin{eqnarray}
V_{0} &=&\frac{\partial \mu _{p}}{\partial \rho
_{p}}-\frac{(\partial \mu
_{n}/\partial \rho _{p})^{2}}{\partial \mu _{n}/\partial \rho _{n}},\text{ }%
k_{TF}^{2}=\frac{4\pi e^{2}}{\partial \mu _{e}/\partial \rho _{e}},  \notag \\
\beta  &=&D_{pp}+2D_{np}\zeta +D_{nn}\zeta ^{2},~~\zeta
=-\frac{\partial \mu _{n}/\partial \rho _{p}}{\partial \mu
_{n}/\partial \rho _{n}},  \notag
\end{eqnarray}%
with $\mu _{i}$ being the chemical potential of particle type $i$.
The three terms in Eq.~(\ref{Vdyn}) represent, respectively, the
contributions from the bulk nuclear matter, the density-gradient
(surface) terms, and the Coulomb interaction. For the coefficients
of density-gradient terms, we use the empirical values of $D_{pp}=D_{nn}=D_{np}=132$ MeV$\cdot $fm$^{5}$,
which are consistent with those from the Skyrme-Hartree-Fock calculations~\cite%
{Oya07,XCLM09}. At $k_{\min }=[(\frac{4\pi e^{2}}{\beta }%
)^{1/2}-k_{TF}^{2}]^{1/2}$, $V_{\rm dyn}(k)$ has the minimal value
of $V_{\rm dyn}(k_{\min })=V_{0}+2(4\pi e^{2}\beta )^{1/2}-\beta
k_{TF}^{2}$~\cite{BPS71,BBP71,Pet95a,Pet95b,Oya07}, and $\rho _{t}$
is then determined from $V_{\rm dyn}(k_{\min })=0$.

In the thermodynamical approach, the stability conditions for the
$npe$ matter are~\cite{Lat07,Kub07,Cal85}
\begin{equation}\label{ther1}
-\left(\frac{\partial P}{\partial v}\right)_\mu>0, \quad{\rm
and}\quad -\left(\frac{\partial \mu}{\partial q_c}\right)_v>0.
\end{equation}
In the above, $P=P_b+P_e$ is the total pressure of the $npe$ matter
with $P_b$ and $P_e$ being the contributions from baryons and
electrons, respectively; $v$ and $q_c$ are the volume and charge per
baryon number; and $\mu=\mu_n-\mu_p$ is the chemical potential of
the $npe$ matter. It can be shown that the first inequality in
Eq.~(\ref{ther1}) is equivalent to requiring a positive bulk term
$V_{0}$ in Eq.~(\ref{Vdyn})~\cite{XCLM09}, while the second
inequality in Eq.~(\ref{ther1}) is almost always satisfied.
Therefore, the thermodynamical stability conditions are simply the
limit of the dynamical one for $k\rightarrow 0$ when the surface
terms and the Coulomb interaction are neglected.

\section{Results for the transition density and pressure}
\label{results}

We have used in our study a momentum-dependent MDI interaction that
is based on the modified finite-range Gogny effective interaction
\cite{Das03}. This interaction,  which has been extensively studied
in our previous work~\cite{LCK08}, is exactly the one used in
analyzing the isospin diffusion data from heavy-ion
reactions~\cite{Che05a,LiBA05c}.

\begin{figure}[tbh]
\begin{center}
\includegraphics[scale=0.8]{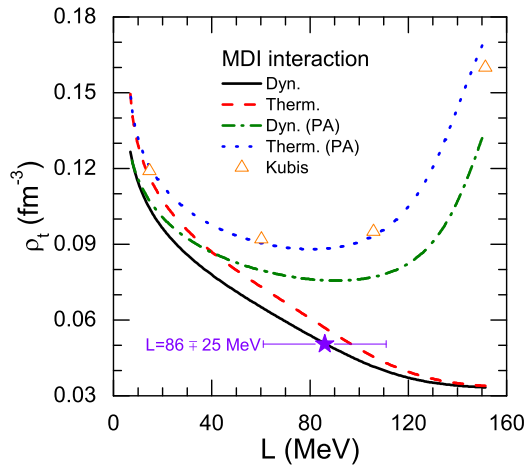}
\caption{{\protect\small (Color online) The transition density
}$\protect\rho _{t}$ {\protect\small as a function of the slope
parameter $L$ of the nuclear symmetry energy from the dynamical and
thermodynamical approaches with and without the parabolic
approximation in the MDI interaction. Taken from
Ref.~\protect\cite{XCLM09}.}} \label{rhotL}
\end{center}
\end{figure}

Shown in Fig.~\ref{rhotL} is the transition density $\rho _{t}$ as a
function of the slope parameter $L$ of the symmetry energy from the
MDI interaction. For comparisons, we have included results from both
the dynamical and thermodynamical approaches with the full EOS and
its parabolic approximation, i.e., $E(\rho ,\delta )=E(\rho ,\delta
=0)+E_{sym}(\rho )\delta ^{2}+O(\delta^{4})$, where
$\delta=(\rho_n-\rho_p)/\rho$ is the isospin asymmetry, from the
same MDI interaction. With the full MDI EOS, $\rho _{t}$ is seen to
decrease almost linearly with increasing $L$ in both approaches
consistent with the results obtained in RPA~\cite{Hor01}. Both
dynamical and thermodynamical approaches give very similar results
with the former having slightly smaller $\rho _{t}$ than the later
(the difference is actually less than $0.01$ fm$^{-3}$), and this is
due to the fact that the density-gradient and Coulomb terms in the
dynamical approach make the system more stable and thus lower the
transition density. The small difference between the two approaches
implies that the effects of density-gradient terms and Coulomb term
are, however, unimportant in determining $\rho _{t}$. On the other
hand, significantly larger transition densities are obtained in the
parabolic approximation, including the predictions by Kubis using
the MDI EOS in the thermodynamical approach~\cite{Kub07}, especially
for stiffer symmetry energies (larger $L$ values). The large error
introduced by the parabolic approach is understandable since the
$\beta $-stable $npe$ matter is usually highly neutron-rich and the
contribution from the higher-order terms in $\delta $ is thus
appreciable. This is especially the case for the stiffer symmetry
energy which generally leads to a more neutron-rich $npe $ matter at
subsaturation densities. Furthermore, because of the energy
curvatures involved in the stability conditions, larger factors are
multiplied to the contributions from higher-order terms in the EOS
than that multiplied to the quadratic term. These features agree
with the early finding \cite{Arp72} that the transition density
$\rho_{t}$ is very sensitive to the fine details of the nuclear EOS.
According to results from the more complete and realistic dynamical
approach, the constrained $L$ limits the transition density to
$0.040$ fm$^{-3} $ $\leq \rho _{t}\leq 0.065$ fm$^{-3}$ as shown in
Fig.~\ref{rhotL}.

\begin{figure}[tbh]
\begin{center}
\includegraphics[scale=0.8]{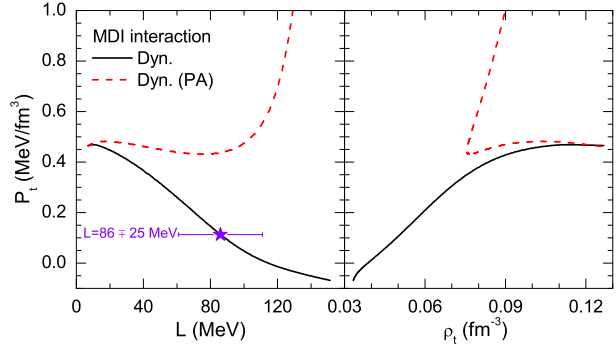}
\caption{{\protect\small (Color online) The transition pressure
$P_{t}$ as a function\protect of $L$ and $ \protect\rho _{t}$ by
using the dynamical approach with and without the parabolic
approximation in the MDI interaction. Taken from
Ref.~\protect\cite{XCLM09}.}} \label{PtLrhot}
\end{center}
\end{figure}

The transition pressure $P_t$ at the inner edge of the neutron star
crust is also an important quantity that might be measurable
indirectly from observations of pulsar glitches \cite{Lat07,Lin99}.
Shown in Fig.~\ref{PtLrhot} is the $P_{t}$ as a function of $L$ and
$\rho _{t}$ by using the dynamical approach with both the full MDI
EOS and its PA. Again, it is seen that the PA leads to huge errors
for large (small) $L$ ($\rho _{t}$) values. For the full MDI EOS,
the $P_{t}$ decreases (increases) with increasing $L$ ($\rho _{t}$)
while it displays a complex relation with $L$ or $\rho _{t}$ in PA.
From the constrained $L$ values, the value of $P_{t}$ is limited
between $0.01$ MeV/fm$^{3}$ and $0.26$ MeV/fm$^{3}$.

\section{Improved constraint on the radius-mass relation of neutron stars}
\label{mr}

The constrained values of $\rho _{t}$ and $P_{t}$ have important
implications in many properties of neutron
stars~\cite{Lat07,Pet95a,Hor04,Oya07}. As an example, we have
examined their effect on constraining the mass-radius ($M$-$R$)
correlation of neutron stars. The crustal fraction of the total
moment of inertia $\Delta I/I$ of a neutron star can be well
approximated by~\cite{Lat00,Lat07,Lin99}
\begin{eqnarray}
\frac{\Delta I}{I} &\approx&\frac{28\pi
P_{t}R^{3}}{3Mc^{2}}\frac{(1-1.67\xi -0.6\xi ^{2})}{\xi } \left[
1+\frac{2P_{t}(1+5\xi -14\xi ^{2})}{\rho _{t}m_{b}c^{2}\xi
^{2}}\right] ^{-1},  \label{dI}
\end{eqnarray}%
where $m_{b}$ is the baryon mass and $\xi =GM/Rc^{2}$ with $G$ being
the gravitational constant. As stressed in Ref.~\cite{Lat00},
$\Delta I/I$ depends sensitively on the symmetry energy at
subsaturation densities through $P_t$ and $\rho_t$, but there is no
explicit dependence on the higher-density EOS. So far, the only
known limit of $\Delta I/I>0.014$ was extracted from studying the
glitches of Vela pulsar \cite{Lin99}. This together with the
upper bounds on $P_t$ and $\rho_t$ ($\rho _{t}=0.065$ fm$^{-3}$ and $%
P_{t}=0.26$ MeV/fm$^{3}$) sets approximately a minimum radius of
$R\geq 4.7+4.0M/M_{\odot }$ km for the Vela pulsar. The radius of
Vela pulsar is predicted to exceed $10.5$ km should it have a mass
of $1.4M_{\odot }$. We notice that a constraint of $R\geq
3.6+3.9M/M_{\odot }$ km for this pulsar has previously been derived
in Ref.~\cite{Lin99} by using $\rho _{t}=0.075$ fm$^{-3}$ and
$P_{t}=0.65$ MeV/fm$^{3}$. However, the constraint obtained in our
study using for the first time data from both the terrestrial
laboratory experiments and astrophysical observations is more
stringent.

\begin{figure}[tbh]
\begin{center}
\includegraphics[scale=0.8]{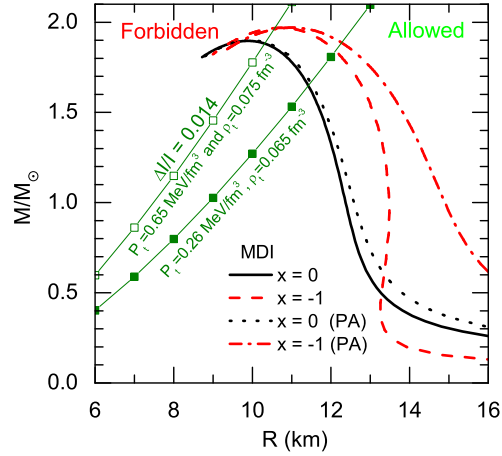}
\vspace{-0.5cm} \caption{(Color online) The mass-radius relation
$M$-$R$ of static neutron stars from the full EOS and its parabolic
approximation in the MDI interaction with $x=0$ and $x=-1$. For the
Vela pulsar, the constraint of $\Delta I/I>0.014$ limits the allowed
masses and radii. See text for details. Taken from
Ref.~\protect\cite{XCLM09} with small modifications.} \label{MR}
\end{center}
\end{figure}

The above constraints are shown in Fig.~\ref{MR} together with the
M$-$R relation obtained by solving the Tolman-Oppenheimer-Volkoff
(TOV) equation. In the latter, we have used the well-known BPS EOS
\cite{BPS71} for the outer crust. In the inner crust with $\rho
_{\rm out}<\rho <\rho _{t},$ the EOS is largely uncertain and
following Ref. \cite{Hor03}, we use an EOS of the form
$P=a+b\epsilon ^{4/3}$ with the constants $a$ and $b$ determined by
the total pressure $P$ and total energy density $\epsilon$ at
$\rho_{\rm out}$ and $\rho _{t}$. The full MDI EOS and its parabolic
approximation with $x=0$ and $x=-1$ are used for the uniform liquid
core with $\rho \geq \rho _{t}$. Assuming that the core consists of
only the $npe$ matter without possible new degrees of freedom or
phase transitions at high densities, the PA leads to a larger radius
for a fixed mass compared to the full MDI EOS. Furthermore, using
the full MDI EOS with $x=0$ and $x=-1$ constrained by the heavy-ion
reaction experiments, the radius of a canonical neutron star of
$1.4M_{\odot }$ is tightly constrained within $11.9$ km to $13.2$
km.

\section{Summery}
\label{summery}

Using the MDI EOS of neutron-rich nuclear matter constrained by
recent isospin diffusion data from heavy-ion reactions in the same
sub-saturation density range as the neutron star crust, we have
determined the density and pressure at the inner edge, that
separates the liquid core from the solid crust of neutron stars, to
be $0.040$ fm$^{-3}$ $\leq \rho _{t}\leq 0.065$ fm$^{-3}$ and $0.01$
MeV/fm$^{3}$ $\leq P_{t}\leq 0.26$ MeV/fm$^{3}$, respectively. These
constraints have allowed us to determine an improved mass-radius
relation for neutron stars. Furthermore, we have found that the
widely used parabolic approximation to the EOS of asymmetric nuclear
matter leads to significantly higher core-crust transition densities
and pressures, especially for stiff nuclear symmetry energies.

\section{Acknowledgements}
\label{acknowledgements}

This work was supported in part by U.S. National Science Foundation
under Grant No. PHY-0758115, PHY-0652548 and PHY-0757839, the Welch
Foundation under Grant No. A-1358, the Research Corporation under
Award No. 7123, the Texas Coordinating Board of Higher Education
Award No. 003565-0004-2007, the National Natural Science Foundation
of China under Grant Nos. 10575071 and 10675082, MOE of China under
project NCET-05-0392, Shanghai Rising-Star Program under Grant No.
06QA14024, the SRF for ROCS, SEM of China, the National Basic
Research Program of China (973 Program) under Contract No.
2007CB815004.

\end{document}